\documentclass[aps,pra,epsfigure,twocolumn,superscriptaddress]{revtex4-1}
\usepackage{dcolumn}    
\usepackage{bm} 
\usepackage{graphicx}
\usepackage{amsmath}    
\usepackage{latexsym}
\usepackage{amsfonts}   
\usepackage{amssymb}
\usepackage{array}      
\usepackage{epsfig}
\usepackage{txfonts}
\usepackage{color}
\usepackage[colorlinks=true,linkcolor=blue,urlcolor=blue,citecolor=blue]{hyperref}
\usepackage{hyperref}
\usepackage{enumerate}

\def \ell{{d}}

\newcommand{\su}{\uparrow}
\newcommand{\sd}{\downarrow}

\newcommand{\Hop}{\hat{H}}

\newcommand{\sop}{\hat{\sigma}}
\newcommand{\Svop}{{\boldsymbol {\hat S}}}
\newcommand{\psiop}{\hat{\Psi}} 
\newcommand{\Bv}{\boldsymbol B}
\newcommand{\Ev}{\boldsymbol E}
\newcommand{\cop}{\hat{c}}
\newcommand{\pop}{\hat{p}}

\newcommand{\rhom}{\hat{\rho}}            
\newcommand{\lan}{\langle}
\newcommand{\ran}{\rangle}

\newcommand{\dagg}{{^{\dagger}}}

\newcommand{\im}{{\rm i}}

\newcommand{\sutd}{Singapore University of Technology and Design, 8 Somapah Road, 487372 Singapore} 
\newcommand{\majulab}{MajuLab, CNRS-UNS-NUS-NTU International Joint Research Unit, UMI 3654, Singapore} 
\newcommand{\belfast}{Centre for Theoretical Atomic, Molecular and Optical Physics, Queen's University Belfast, Belfast BT7 1NN, United Kingdom} 

\newcommand{\ket}[1]{\left\vert#1\right\rangle}
\newcommand{\bra}[1]{\left\langle#1\right\vert}

\begin{document}
\title{Cost of counterdiabatic driving and work output}
\author{Yuanjian Zheng} 
\affiliation{\sutd}
\author{Steve Campbell} 
\affiliation{\belfast}
\author{Gabriele De Chiara} 
\affiliation{\belfast}
\author{Dario Poletti} 
\affiliation{\sutd} 
\affiliation{\majulab}

\begin{abstract}
Unitary processes allow for the transfer of work to and from Hamiltonian systems. However, to achieve non-zero power for the practical extraction of work, these processes must be performed within a finite-time, which inevitably induces excitations in the system. We show that depending on the time-scale of the process and the physical realization of the external driving employed, the use of counterdiabatic quantum driving to extract more work is not always effective. We also show that by virtue of the two-time energy measurement definition of quantum work, the cost of counterdiabatic driving can be significantly reduced by selecting a restricted form of the driving Hamiltonian that depends on the outcome of the first energy measurement. Lastly, we introduce a measure, the exigency, that quantifies the need for an external driving to preserve quantum adiabaticity which does not require knowledge of the explicit form of the  counterdiabatic drivings, and can thus always be computed. We apply our analysis to systems ranging from a two-level Landau-Zener problem to many-body problems, namely the quantum Ising and Lipkin-Meshkov-Glick models.  
\end{abstract}
\pacs{03.67.Ac, 05.70.Ln, 75.10.Jm, 37.10.Jk} 

\maketitle

\section{Introduction}
Recent years have witnessed a surge of interest in the study of thermal nanomachines that are capable of converting disordered forms of energy, such as heat, into useful work. At such small scales,  thermal and quantum fluctuations play a considerable role, and as such the work output and performance of an engine are characterized probabilistically by distribution functions.  
These distributions obey fluctuation theorems such as the Jarzynski equality and the Crooks equation~\cite{Jarzynski1997,Crooks1999,CampisiTalkner2011,HanggiTalkner2015},
which have been verified by experiments both at the classical~\cite{WangEvans2002,ImparatoSasso2007,GomezSolanoCiliberto2010,BlickleBechinger2006,ImparatoCiliberto2008,HummerSzabo2001,LiphardtBustamante2002,CollingBustamante2005,GuptaWoodside2011,DouarcheCiliberto2006,GarnierCiliberto2005,SairaPekola2012,GomezSolanoBechinger2015} and quantum level~\cite{PaternostroPRL,AnKim2015}. There have also been recent implementations of miniature classical engines \cite{SteenekenVanBeek2011, BlickeBechinger2012, LeeOzyilmaz2014, MartinezRica2014, RossnagelSinger2016} and several proposals for the realization of quantum heat engines \cite{HallwoodPRL, AbahLutz2012, ZhangMeystre2014, CampisiFazio2015}. Effects of quantum statistics of the working fluid have also been investigated \cite{YiTalkner2012, GongQuan2014, ZhengPoletti2015}.           

On the other hand, it has been shown for both classical and quantum systems that external drivings can allow a system to evolve adiabatically even when driven in finite time \cite{DemirplackRice2008, Berry2009, DengGong2013, PalmeroMuga2013, TorronteguiMuga2013, Jarzynski2013, DeffnerDelCampo2014, DelCampoPaternostro2014, DelCampo2013, AcconciaDeffner2015, DelCampoZurek2012, CampbellFazio2014, RohringerTrupke2015}. This has applications in quantum control, and can be performed in three ways: (i) Driving of a system such that, instantaneously, a state evolves adiabatically (e.g. \cite{DemirplackRice2008, Berry2009, DengGong2013, PalmeroMuga2013, TorronteguiMuga2013, Jarzynski2013, DeffnerDelCampo2014, DelCampoPaternostro2014}). This is known as counterdiabatic driving and as transitionless driving; 
(ii) Protocols for which only at the final time are the states adiabatically transferred, while there may be excitations at intermediate times of the process (e.g. \cite{DelCampo2013, AcconciaDeffner2015}); (iii) Application of imperfect external drivings which do not allow for an exact adiabatic transfer, but close enough for most practical purposes (e.g. \cite{DelCampoZurek2012, CampbellFazio2014, RohringerTrupke2015}).         

Hence it was suggested \cite{DelCampoPaternostro2014} to use such external drivings to render the unitary processes of a thermodynamic cycle quantum adiabatic while being performed in finite time. This could considerably augment the performance of nano-thermodynamic engines as work exchanges are extremized by adiabatic protocols \cite{AllahverdyanNieuwenhuizen}. However, implementing additional external driving requires resources which affect the overall performance of the system \cite{MugaarXiv}.         

In this paper we analyze the implications of considering the necessary power in applying counterdiabatic driving both in a prototypical system such as the Landau-Zener model \cite{Landau1932, Zener1932, Stueckelberg1932, Majorana1932} and also in many-body quantum systems such as the transverse field Ising chain \cite{Lenz1920, Ising1925}. 
Subsequently, we then show that this cost may outweigh the possible gains in work extraction for slow enough processes due to the relative degree of adiabaticity in the dynamics. Conversely, for relatively faster processes, the use of counterdiabatic driving can improve the work exchange, depending on the experimental realization of the fields. Furthermore, 
we devise a general strategy that exploits the definition of work as a two-time measurement of energy \cite{TalknerHanggi2007} to improve the performance of work transfer. In particular, we show that it is possible to achieve sizable energy savings by gathering information from the first measurement and then applying a specifically tailored driving to the protocol.  Lastly, we introduce an alternative measure, the exigency, to quantify the need for applying couterdiabatic driving which is related to the non-commutativity between the time derivatives of the Hamiltonian and the state. This measure has the advantage in that it can always be computed regardless of whether the protocol for counterdiabatic driving is known. Moreover, it mimics the behavior of the cost functions associated to the transitionless form of counterdiabatic driving and goes to zero when no external driving is needed. We apply this measure to the analysis of the quantum harmonic oscillator and the Lipkin-Meshkov-Glick infinite range spin model.
\section{Counterdiabatic Driving}
\subsection{The Transitionless Protocol}
To begin with, we consider the evolution of a density matrix $\rhom(t)$ from $t_0$ to $t_1$ under unitary dynamics of the Hamiltonian $\Hop_0(t)$. Counterdiabatic quantum driving is obtained by applying an external Hamiltonian 
\begin{align} 
\Hop_{t}(t)=\im\hbar\sum_j \frac{d\hat{P}_j(t)}{dt}\hat{P}_j(t),
\label{eq:Htl}      
\end{align}  
where $\hat{P}_j=|j\rangle\langle j|$ is the projection operator on the instantaneous energy eigenstate $|j\rangle$ of $\Hop_0=\sum_j E_j(t) |j\rangle\langle j|$ \cite{DemirplackRice2008, Berry2009, Kato1950, Messiah1965, notunique}.  
     

{ Note that the driving in Eq.(\ref{eq:Htl}) is such that all energy eigenstates evolve transitionlessly. Such a strong requirement is however not necessary when using counterdiabatic driving to enhance the work output. The mean work exchange of a single unitary process is defined by a two time measurement protocol $\langle W\rangle_{t_0\rightarrow t_1} = \sum_{m,n} \left[E_m(t_1)-E_n(t_0)\right]\mathcal{P}^{m,n}p_n(t_0)$ where $E_m$ is the instantaneous energy of the $m$-th level, $p_n(t_0)$ is the probability of occupying that level at $t_0$, and $\mathcal{P}^{m,n}$ is the transition probability from level $n$ at time $t_0$ to level $m$ at time $t_1$ \cite{TalknerHanggi2007}. 

Next, as consequence of a resolved energy measurement performed before the unitary process, the system may collapse to one of its instantaneous energy eigenstates. 
Hence, given a particular outcome of the first energy measurement, we can apply a suitable driving which preserves the transitionless evolution of only the measured state (or a relevant sub-manifold of the entire system) without the need of avoiding transitions between the other levels that remain unpopulated throughout. Such a selected counterdiabatic driving for a particular eigenstate $j$ is given by \cite{DemirplackRice2008} }              
\begin{align}
\Hop_{_W,j}&=\im \hbar \left[\frac{d\hat{P}_j}{dt},\hat{P}_j\right].
\label{eq:Hs} 
\end{align} 

{
Here we point out that while the present discussion is primarily in context of unitary evolution, which is important in its own right, it is also relevant to the more general context of quantum engine cycles that consists of different strokes. For instance an Otto cycle is composed of two unitary strokes and two strokes in which only heat is transferred with the environment. An initial density operator that describes the system, under the (sequential) repetitive application of these strokes, generally reaches an asymptotic dynamic behavior which can be used to characterize the cycle. However, characterization of the performance of such an engine cycle includes determining its net work done and/or efficiency. One approach to achieve this is to perform energy measurements after each stroke (however, we remark that in principle one could also determine the work done using indirect probes\cite{DornerVedral2013,MazzolaPaternostro2013,DeChiaraPaz2015}). These measurements would then in turn affect the dynamics of the system changing the asymptotic behavior of the dynamical evolution. It is thus important to also consider the energy measurements already within the cycle. See \cite{ZhengPoletti2016, HayashiTajima2015} for a more in-depth discussion of measurement within the strokes of an engine cycle. 

} 

\subsection{Power in Generating Counterdiabatic Fields }
To quantify the power required to generate such external driving it is instructive to consider two examples: 

\begin{enumerate}[i.]
\item A single spin in a time-dependent magnetic field ${\Bv_0}(t)$, with Hamiltonian $\Hop_{0,s}=\gamma\Bv_{0}(t) \cdot \Svop$. Here, $\gamma$ is the gyromagnetic ratio and $\Svop$, for a spin-$1/2$ system, is given by $\Svop=\left(\hbar/2\right){\boldsymbol \sop}$ where ${\boldsymbol \sop}$ is the vector composed of the Pauli matrices. 
\item A neutral atom in a time-dependent electric field $\Ev(t)$ such that it experiences a potential $V(t)\propto |\Ev(t)|^2$.
\end{enumerate}

In (i) the part of the Hamiltonian attributed to the applied field is given by $\Hop_{b}= \Bv_1 \cdot \Svop$ where $\Bv_1=\frac{1}{B_0^2}\Bv_0 \times \left(\frac{\partial \Bv_0}{\partial t}\right)$ \cite{Berry2009} which can be generated by an electric current $I(t)$ such that the power required would scale as $|\Bv_1|^2$ and thus be proportional to $\|\Hop_b\|^2$. However for (ii), this term could instead be proportional to the modulus square of another electric field $V_E(t) \propto |\Ev'(t)|^2$, and thus the power needed would then scale with the norm of the driving but not the square of it as in the previous case. 

Thus, while the power required to generate the counterdiabatic drivings scales as the norm of the driving Hamiltonian (we use the Frobenius norm $\|\hat{A}\|=\sqrt{\text{Tr}\left[\hat{A}^{\dagger}\hat{A}\right]}$ where $\hat{A}$ is an operator), the exact functional dependence on power is strongly affected by the experimental realization \cite{MagneticVsElectric}. Hence, the cost of counterdiabatic driving can be written in general as 
\begin{align}
C_{t}^n&= \nu_{t,n}\int_{t_0}^{t_1} \|\Hop_{t}\|^n dt,
\label{eq:C1} 
\end{align} 
where $\nu_{t,n}$ is a set-up dependent constant and the index of the norm $n$ depends on the nature of the applied fields \cite{MagneticVsElectric, SameN}.
The principle of its usage here as a measure of cost is similar to the constraints used in optimal quantum control studies \cite{PeirceRabitz1988, KosloffTannor1989}. A closely related measure has also recently been used in the context of energetic cost of superadiabatic computations~\cite{santos}.
%
It follows that we can define the cost of applying the selected counterdiabatic Hamiltonian $\Hop_W$ in a similar fashion:
\begin{align}
C_{_W}^n&= \sum_j \nu_{_W,j,n}~p_j \int_{t_0}^{t_1} \|\Hop_{_W,j}\|^n dt \;,
\label{eq:Cs} 
\end{align} 
which can be interpreted as the weighted average of the cost of driving each level $j$, over the level occupation probability $p_j={\rm tr}(\rhom \hat{P}_j)$. In fact the frequency of use of a particular driving depends on the probability of measuring that particular energy level. Similar to $\nu_{t,n}$, $\nu_{_W,j,n}$ is a parameter dependent on the particular experimental set-up, and in the following we set $\nu_{t,n}=\nu_{_W,j,n}=1$ for simplicity. { Here we stress that while (\ref{eq:C1}) and (\ref{eq:Cs}) are entirely general expressions applicable to counterdiabatic fields, the exact functional dependence {\bf ($n$)} and energy scale {\bf $\nu_{t,n}$ or $\nu_{_W,j,n}$} are dependent on the particular form of the driving and physical nature of the fields which, as demonstrated by the two examples considered, cannot be generalized.} 

We are now equipped to analyze the implications of considering the cost of counterdiabatic driving in the performance of a work protocol. In the following, we focus on the Landau-Zener model followed by the Ising model for which the counterdiabatic driving term is known analytically and is shown to be closely related to that of the Landau-Zener model \cite{DelCampoZurek2012}. 


\section{Landau-Zener Model}
{ A prototypical model for the study of quantum dynamics is the Landau-Zener model \cite{Landau1932, Zener1932, Stueckelberg1932, Majorana1932} which in some cases, can even allow for analytical insights into the dynamics. It consists of a two-level system with the Hamiltonian}:
\begin{equation}
\Hop_{LZ}=g(t)\hat{\sigma}^z+\Delta\hat{\sigma}^x \label{eq:LZmodel}
\end{equation}
{ The Landau-Zener model describes the dynamics due to a time-dependent modulation of $g(t)$ through an avoided level crossing with a finite energy gap $\Delta$. For driving protocols that are not infinitely slow (i.e the time derivative $\dot{g}(t) \neq 0$), the transition probability between the two levels becomes non-zero and the application of a counterdiabatic field is required to recover transitionless dynamics \cite{Berry2009,DelCampoZurek2012}. Hence Eq.~(\ref{eq:Htl}) for the Landau-Zener model (\ref{eq:LZmodel}) becomes}  
\begin{equation}
\Hop_{LZ,t}=-\frac{\hbar \dot{g}(t)\Delta}{2(\Delta^2+g^2)}\hat{\sigma}^y \label{eq:LZtrans}
\end{equation}
Note that it being a two-level system, the driving in Eq.~(\ref{eq:Htl}) $\Hop_{LZ,t}$ is identical to $\Hop_{LZ,_W}$ from Eq.~(\ref{eq:Hs}) \cite{2levels}. We now compare the work done in absence of any additional external driving $\lan W\ran$, with that done for an adiabatic process $\lan W_{ad}\ran$ considered in conjunction with the cost of counterdiabatic driving. The counterdiabatic driving is beneficial only when the cost of producing it is lesser than the inner friction $\lan W_{fric}\ran=\lan W\ran - \lan W_{ad} \ran$ generated in absence of the counterdiabatic fields which incidentally, is also the amount of additional work extractable attributed to the driving.      	

For a Landau-Zener process where the state is initially far enough from the avoided crossing, the probability of populaton transfers decays exponentially with the timescale. This implies that the inner friction $\lan W_{fric}\ran \propto \exp[-\alpha (t_1-t_0)]$ also decays exponentially, where $\alpha$ is time-independent. As such, the cost of the driving would need to decay at least exponentially fast with the increase of the time scale $t_1-t_0$,  or the cost of driving would at some point be greater than the gains obtained by a perfect adiabatic evolution. However, we find that the cost of driving always decays as a power law:
\begin{equation}
C_t^{n}\propto \frac 1 {(t_1-t_0)^{n-1}}
\end{equation}
This can be shown by a simple change of variables $s=(t-t_0)/(t_1-t_0)$ that yields $C_{t}^n=\left[1/(t_1-t_0)^{n-1} \right]\int_0^1 \|\Hop_t(s)\|^n ds$ ($C_{_W}^n$ behaves analogously). It should be noted that this scaling is completely independent of the system and protocol used. Hence, the usefulness of counterdiabatic driving will ultimately always depend on the particular process in question.

Despite this inherently system specific nature of the relative behaviors of cost and inner friction, we introduce an entirely general strategy that reduces the cost of counterdiabatic driving regardless of the particular experimental realization. This involves noticing that while for a single two level system $\Hop_t$ is identical to $\Hop_{_W,j}$, and would thus cost the same amount, they are vastly different for larger systems, and the resulting difference in their cost can be very significant. 

\section{Ising Model}

To illustrate this, we focus on the transverse field Ising model because it allows for greater analytical insights with the exact form of the counterdiabatic field known, while at the same time presenting a phase transition. Its Hamiltonian reads   
\begin{align} 
\Hop_{I} = -\sum_{i}  J\sop_{i}^x\sop_{i+1}^x + g(t)\sop_{i}^z. 
\label{eq:Isingspin}   
\end{align} 
In Eq.~(\ref{eq:Isingspin}), $J$ is the amplitude of spin excitation tunneling while $g(t)$ is a time-dependent transverse magnetic field. In the following we use a smooth ramp 
\begin{align}
g(t)=g_0+(g_1-g_0)\{1-\cos[\pi (t-t_0)/(t_1-t_0)]\}/2 \label{eq:goft}
\end{align}   
in order to avoid sudden quenches \cite{DelCampo2013}.  

\begin{figure}
\includegraphics[width=\columnwidth]{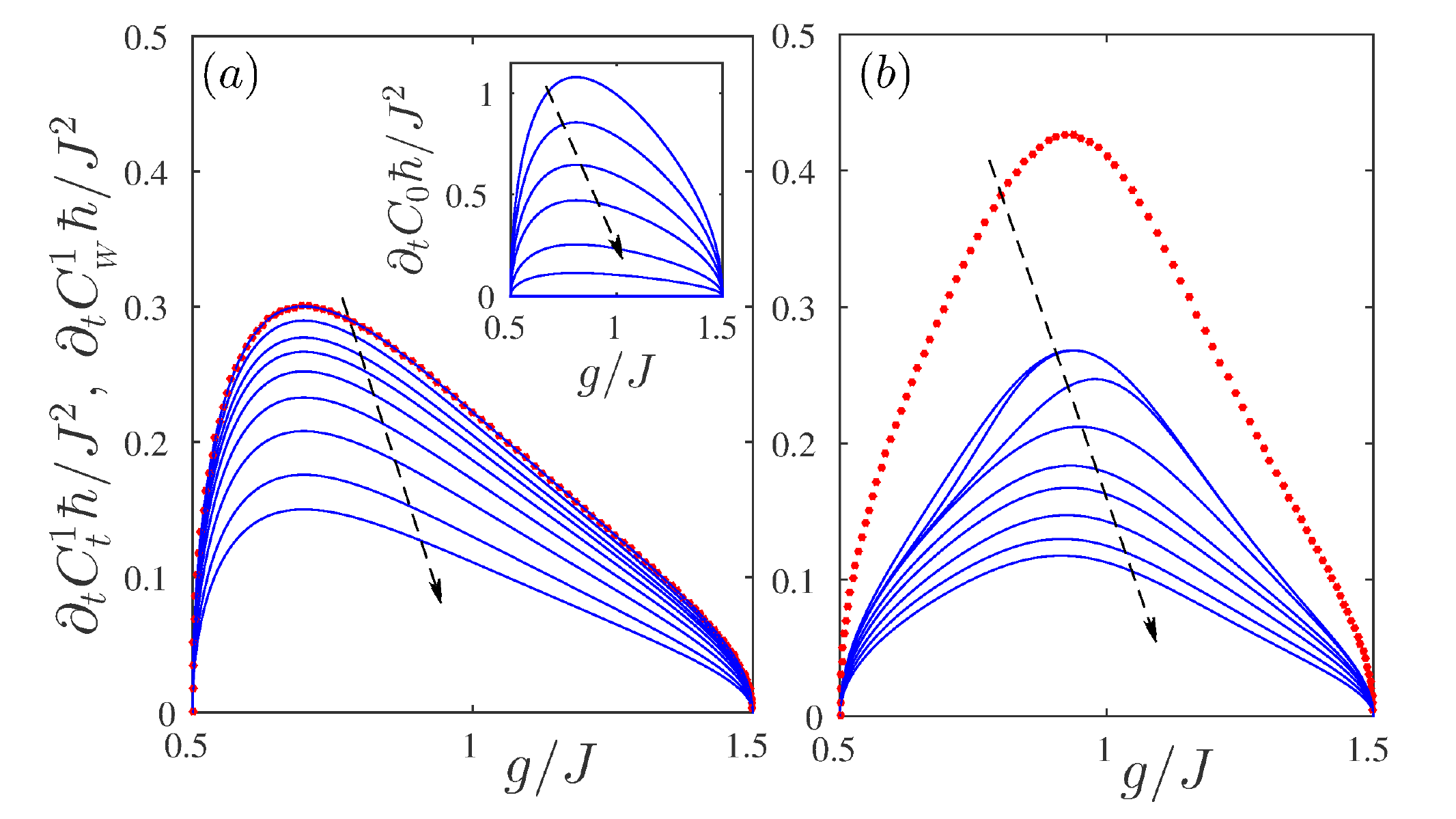} 
\caption{(Color online) (a,b) Instantaneous cost of counterdiabatic driving for two spins (a) and $8$ spins (b). The red circles represent $\partial_tC_t^1$ while blue continuous lines depict $\partial_tC_{_W}^1$ for decreasing values of the inverse temperature (as indicated by the arrow) $\beta=\infty$ (top) to $\beta=0$ (bottom). The inset of (a) shows the instantaneous exigency $\partial_t C_0$ also from $\beta=\infty$ (top) to $\beta=0$ (bottom).}
\label{fig:1} 
\end{figure}

Now, we consider the case of two spins revealing the basic principles of our strategy. The Hamiltonian, which we refer to as $\Hop_{2s}$, is divided into two blocks, one which dynamically couples the state with two spin-ups $\lvert\su\su\ran$ with $\lvert \sd\sd\ran$, and the other which couples $\lvert \su\sd\ran$ with $\lvert \sd\su\ran$. { The first block reverts to a two-level Landau-Zener problem, which, as in (\ref{eq:LZtrans}), requires the application of the counterdiabatic Hamiltonian}  
\begin{equation}
\Hop_{I2,t}=-\frac{\im \hbar \dot{g}(t)}{J^2+4g^2(t)} \left\{\lvert \su \su \rangle\langle \sd \sd \rvert - \lvert \sd \sd \rangle\langle \su \su \rvert \right\}. 
\end{equation}
The second block is time-independent and hence requires no external driving. Considering a thermal state as the initial condition and a change of $g$ from $0.5J$ to $1.5J$ we observe a significant difference in $C_t^n$ and $C_{_W}^n$ (Fig.\ref{fig:1}). In fact with a probability that is dependent on temperature, the first energy measurement could pick a state in the undriven sector which would thus require no external driving such that $C_{_W}^n\le C_t^n$. In particular $C_{_W}^n$ can be as small as $C_t^n/2$ for small $\beta$.                              
\begin{figure}
\includegraphics[width=\columnwidth]{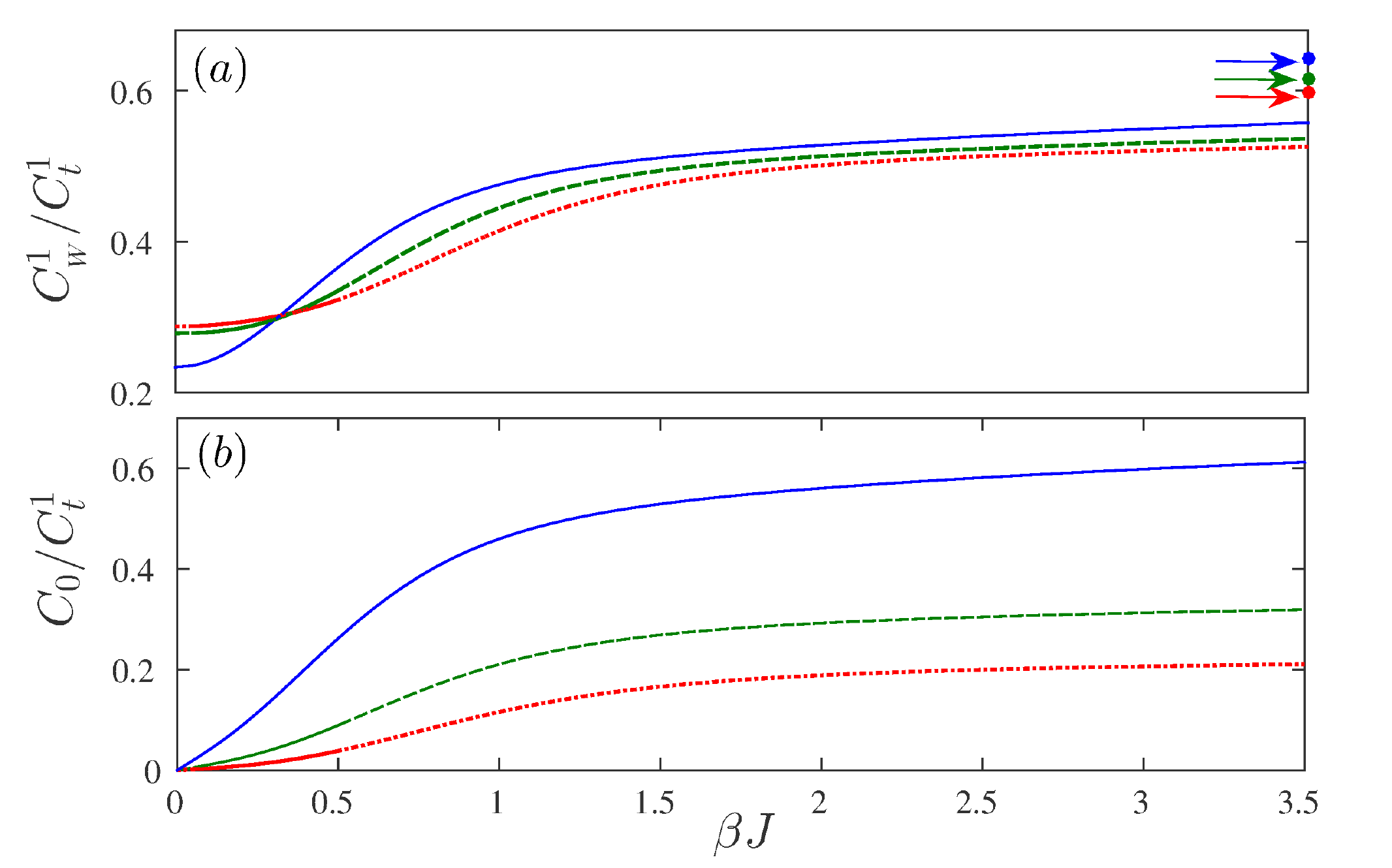} 
\caption{(Color online) (a) Ratio of costs $C^1_W/ C^1_t$ versus inverse temperature $\beta$. The arrows indicate the asymptotic value computed for the groundstate (at $\beta=\infty$). (b) $C_0 / C_t^1$ versus $\beta$. The blue continuous line represents $L=4$, green dashed line $L=6$ and red dotted line $L=8$.}
\label{fig:1b} 
\end{figure}
%

When considering a larger system the many-body physics makes this study even richer. The Ising model in Eq.~(\ref{eq:Isingspin}) exhibits a quantum phase transition at $g=\pm J$ in the thermodynamic limit, with the groundstate being paramagnetic for $|g|>J$ and ferromagnetic for $|g|<J$. Here $g(t)$ is varied in time in order to cross $|g|=J$, without considering the case $N\rightarrow\infty$ because no counterdiabatic driving can be exactly done in that regime \cite{DelCampoZurek2012}. For $L$ spins (which we take to be even in the following), we use the Jordan-Wigner transformations \cite{JordanWigner1928, LiebMattis1961, Dziarmaga2005} $\sop_{i}^z=1-2\cop_i^{\dagger}\cop_i$ and $\sop_{i}^x=-\left( \cop_i + \cop^{\dagger}_i\right)\prod_{j<i}\left(  1-2\cop^{\dagger}\cop_j \right) $ on $\Hop_{I,s}$ to write Eq.~(\ref{eq:Isingspin}) in the free-fermion form in momentum space
\begin{align} 
\Hop_{I,k} =& \hat{P}_e \left(\!\sum_{\;k\in k_e}\Hop_{k} \!\right)\hat{P}_e +  \hat{P}_o \left( \sum_{k\in k_o}\Hop_{k} + \Hop_{0,\pi} \right)  \hat{P}_o,
\label{eq:Isingfermi}
\end{align} 
where $\Hop_k=\psiop_k\dagg \left\{ \left[g(t)-J\cos(k)\right]\sigma^z - J\sin(k)\sigma^x \right\}\psiop_k$, $\Hop_{0,\pi}=2g(1-\cop_0\dagg\cop_0-\cop_{\pi}\dagg\cop_{\pi})$. $\psiop_k\dagg = \left(\cop_k\dagg,\;\cop_{-k}\right)$ and $\hat{P}_e$ $\left(\hat{P}_o\right)$ projects on the even (odd) sector corresponding to the space containing an even (odd) number of fermionic excitations. In the even sector, due to anti-periodic boundary conditions, $k$ takes the values $k_e=\pm (2j-1)\pi/L$ for $j\in [1,L/2]$ while in the odd sector, with periodic boundary conditions, $k_o=\pm 2j\pi/L$ for $j\in [1,L/2)$ in addition to $k_o=0$ and $\pi$~\cite{LiebMattis1961,Dziarmaga2005}. As shown in~\cite{DelCampoZurek2012}, where the authors are concerned with the groundstate, and focused on only the even sector, $\Hop_{I,k}$ is a sum of independent Landau-Zener transitions and the counterdiabatic driving $\Hop_{I,t}$ is thus given by
\begin{align} 
\Hop_{I,t} = \sum_{0<k<\pi} f(k,t)\psiop_k\dagg\sigma^y_k\psiop_k,
\label{eq:H1Ising} 
\end{align} 
where $f(k,t)=-\hbar \dot{g}(t)J\sin(k)/\left[2\left(g^2+J^2-2g J \cos(k)\right)\right]$ and $\psiop\dagg_k=\left(\cop\dagg_k,\cop_{-k}\right)$.  Again, $\Hop_{I,_W}$ depends on the state(s) selected upon the first energy measurement. Thus, in a similar fashion to the case of the two-spins, (albeit in a far richer way) the first energy measurement may select a reduced portion of the even or odd sectors: $\Hop_{I,_W}=\hat{P}_{\{\tilde{k}\}}\Hop_{I,t}\hat{P}_{\{\tilde{k}\}}$, where $\hat{P}_{\{\tilde{k}\}}$ projects over only the relevant set of quasi-momenta $\tilde{k}$ (e.g. $\hat{P}_{\{k\}}$ projects only over the set of even eigenstates), that are connected by pair creation or destruction operators (see Appendix A). As a result $C_{I,_W}/C_{I,t}$ can be considerably reduced especially at high temperatures [see Fig.~\ref{fig:1b} (a)] \cite{identnu}.                 
The reduction in the energy cost of the counterdiabatic driving can be further reduced for longer spin chains and is more significant when the initial condition does not include states whose dynamics entails small avoided crossings (in this case the cost could be negligible). 

{ Our results show that it becomes more costly to drive the Ising model transitionlessly as $\beta \to \infty$ (i.e. at zero temperature). This might seem counter-intuitive since the system increasingly approaches its groundstate and thus would only require a single eigenstate to be driven. However, while in general there can be considerable energy savings when the first energy measurement selects only a single state, the cost of guaranteeing adiabatic dynamics may become considerably large when this particular state happens to be the groundstate and that the system is going across a quantum phase transition. In fact, it would cost an infinite amount of energy (for an infinitely large system) to evolve in a perfectly adiabatic manner \cite{DelCampoZurek2012, CampbellFazio2014}. As we consider finite systems, the corresponding cost for driving the groundstate will not be infinite, but instead be typically larger.

}
\section{Exigency}
While $C_t^n$ and $C_{_W}^n$ indicate the energy cost of achieving quantum adiabatic dynamics with external drivings, it cannot always be computed because it requires the knowledge of the counterdiabatic field. Moreover the counterdiabatic driving always ensures absence of transitions even when they would be perfectly balanced. For example the identity matrix commutes with any Hamiltonian hence it does not require any driving to preserve it. We thus look for an indicator for the need of using counterdiabatic driving which (i) would be non-zero when a driving is needed and (ii) can always be computed. 
We thus study the origin of the need of counterdiabatic driving by analyzing the evolution of a density matrix which is given by $\rhom(t_1)=\hat{U}\rhom(t_0)\hat{U}^{\dagger}$ with $\hat{U}=\mathcal{T}{\rm exp}[-{\im}\int_{t_0}^{t_1}\Hop_0(t)dt/\hbar]$, where $\mathcal{T}$ stands for the time ordering operator. Since the initial conditions considered are diagonal in the basis of the initial Hamiltonian $\Hop_0(t_0)$ (this includes all thermal states), to the lowest order in $dt$ the evolution of $\rhom$ is given by    
\begin{align}
\rhom(t+dt)=& \rhom(t) -\frac{{\im}}{\hbar}\frac{dt^{n+1}}{n!}\left[ \Hop_0^{(n)}(t), \rhom(t) \right] +O(dt^{n+2}),   
\label{eq:evolution}     
\end{align}  
where $\Hop_0^{(n)}(t)$ is the $n$-th derivative of $\Hop_0(t)$ and $n$ is the lowest natural number for which the commutator in Eq.~(\ref{eq:evolution}) is non-zero, details can be found in Appendix B. 
This implies that the first term of Eq.~\eqref{eq:evolution} which may contribute is at least of second order in $dt$. It follows that the instantaneous power dissipated by internal friction is $\mathcal{P}=\lim_{dt\rightarrow 0} \lan \delta W_{fric}\ran/dt=0$ \cite{FeldmannKosloff2000, KosloffFeldmann2002, KosloffFeldmann2003, FeldmannKosloff2004, FeldmannKosloff2012, WangMa2013, PlastinaZambrini2014}, where $\lan \delta W_{fric}\ran=\langle \delta W\rangle - \langle \delta W_{ad}\rangle$ and $\langle \delta W\rangle$ is the actual infinitesimal work while $\langle\delta W_{ad}\rangle$ is the infinitesimal work done if the process was quantum adiabatic. This however does not mean that applying an external driving to make the evolution quantum adiabatic requires no power. The fact that the instantaneous variation of inner friction over time $\lan \delta W_{fric} \ran/\delta t$ is negligible for continuous drivings implies that it is not the best measure of neither the need of counterdiabatic driving nor of its cost. Note that this is different from Ref.~\cite{VacantiAuffeves2015} where the density matrix instantaneously does not commute with the Hamiltonian.

\begin{figure}
\includegraphics[width=0.9\columnwidth]{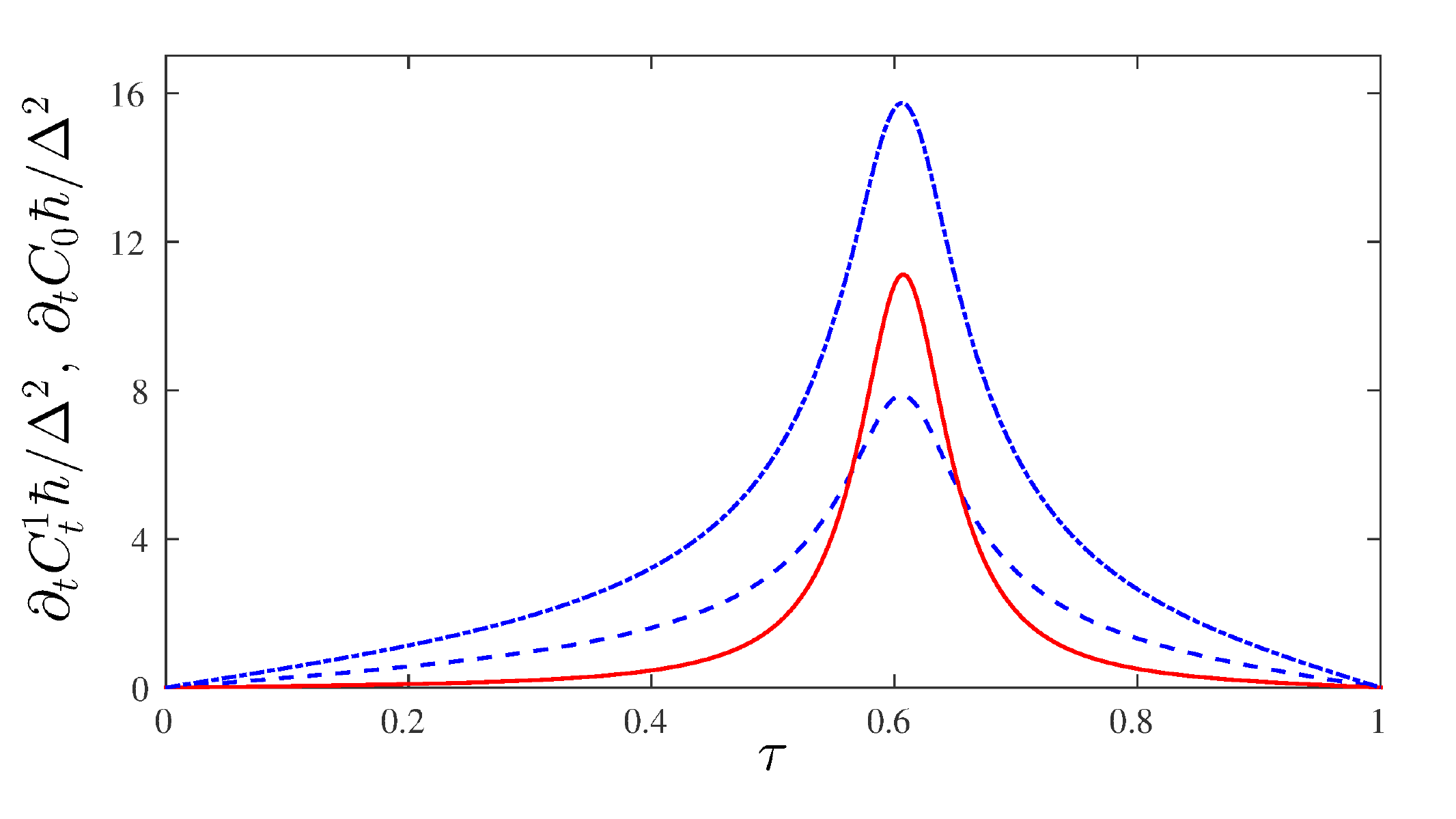}
\caption{(Color online)  Comparison of Exigency to the cost of transitionless driving in the Landau-Zener model [Eq.(\ref{eq:LZmodel})] as $g(t)$ is driven from $g(t_0)=-10\Delta$ to $g(t_1)=5\Delta$ following also Eq.(\ref{eq:goft}). Red continuous line represents $\partial_t C_t^1$ while the other lines show $\partial_t C_0$  for an initial groundstate occupation of $p_g=1$ (blue dot-dashed line) and $p_g=0.75$ (light blue dashed line). Here $\tau=(t-t_0)/(t_1-t_0)$ and $(t_1-t_0)=\Delta^{-1}\hbar$.}
\label{fig:2} 
\end{figure}

From Eq.~(\ref{eq:evolution}), we measure the need for counterdiabatic driving using the quantity $C_0$, which we refer to as exigency, 
\begin{align}
C_{0}= \int_{t_0}^{t_1}\left\| \left[ \frac{\partial \Hop_0(t)}{\partial t}, \rhom(t)   \right] \right\| dt .   
\label{eq:C0}    
\end{align}
Eq.~\eqref{eq:C0} provides a qualitative understanding of the cost of driving, making it remarkably useful as it can be readily calculated for any Hamiltonian. { $C_0$ measure the degree of non-commutativity between $\rhom$ and the Hamiltonian which, if non-zero, implies the need of counterdiabatic driving}. Moreover, (i) $C_0 \to 0$ as $\beta\rightarrow 0$ as desired, since $\rhom$ is proportional to the identity and commutes with any time-dependent Hamiltonian and thus not need any counterdiabatic driving, see inset of Fig.\ref{fig:1} (a)  and Fig.\ref{fig:1b} (b), and (ii) the instantaneous cost $\partial_t C_0$, as shown in the inset of Fig.\ref{fig:1} (a) and in Fig.\ref{fig:2}, mimics that of the counterdiabatic driving $\partial_t C^1_t$ { and similarly for larger $n$ in $C^n_t$}.   
%
In particular in Fig.~\ref{fig:2} we plot the instantaneous cost of the external driving $\partial_t C_{t}^n$ for a Landau-Zener problem described by $\Hop_{LZ}$  with the same $g(t)$ of Fig.~\ref{fig:1} and compare it to the instantaneous power estimated from Eq.~(\ref{eq:C0}), i.e. $\partial_t C_0$. While these different measures cannot be exactly compared to each other because of the different, experimentally determined, constant $\nu_{\mu,n}$, they behave similarly. This means that, even without knowing the exact form of the counterdiabatic driving term, $\Hop_{t}$, it is still possible to have a qualitative understanding of the cost of the driving by studying $C_0$. We should also note, however, that the maximum instantaneous cost is not always maximum at a minimum of distance between energy levels in avoided crossings (as in the standard Landau-Zener problem) because it also depends on the exact time-dependence of the Hamiltonian parameters. For more asymmetric cases the instantaneous exigency $\partial_t C_0$ can differ, even qualitatively, from the cost function $\partial_t C_t^n$. To further illustrate the advantage of the exigency we conclude with some additional examples in the next subsections: the harmonic oscillator and the Lipkin-Meshkov-Glick model.  

\subsection{Exigency: Harmonic Oscillator}
The Hamiltonian for the quantum harmonic oscillator with time dependent frequency $\omega(t)$ is given by 
\begin{equation}
\hat{H}^\text{ho}_0= \frac{\hat{p}^2}{2m} + \frac{m}{2}[\omega(t)]^2 \hat{x}^2 ,
\end{equation}
where $m$ is the mass and $\hat{p}$ is the momentum operator. 
In Ref.~\cite{mugaJPB} the exact counterdiabatic term was found to take the simple form
\begin{equation}
\hat{H}^\text{ho}_t = -\frac{\dot{\omega}}{4\omega} \left( \hat{x}\hat{p}+\hat{p}\hat{x} \right).
\end{equation}
where `$\dot{~~}$' refers to the time derivative. Since $\hat{H}^\text{ho}_0$ is unbounded  $C^1_t$ is not finite. However the exigency can still be used to establish the need to perform the counterdiabatic driving. In particular, we find the instantaneous power of the harmonic oscillator, as estimated by Eq.~\eqref{eq:C0} for a given eigenstate $\ket{\psi}$, reduces to
\begin{equation}
\label{costH0}
\partial_t\mathcal{C}_{0} = \sqrt{2} m \vert \dot{\omega} \vert \omega \sqrt{ \big< \psi \vert \hat{x}^4 | \psi \big> - \big< \psi \vert \hat{x}^2 | \psi \big>^2 }.
\end{equation}

For the simple but indicative case of the groundstate this expression can be easily evaluated giving, $\psi(x) = \bra{ x } \psi\big> = \left(\frac{m\omega}{\pi\hbar}\right)^{1/4} \exp\left(-\frac{m\omega x^2}{2\hbar}\right)$, and find $\big< \psi \vert \hat{x}^4 | \psi \big>  = 3\hbar^2/(4m^2\omega^2)$ and $\big< \psi \vert \hat{x}^2 | \psi \big>  =  \hbar/(2m\omega)$. Substituting these expressions into \eqref{costH0} we finally arrive at
\begin{equation}
\label{costH02}
\partial_t C_{0} = \hbar\left| \dot{\omega} \right|.
\end{equation}
If we employ a similar ramp to that used previously,  $\omega(t) = \omega_0 + \frac{\omega_1-\omega_0}{2}\left( 1-\cos\left(\frac{\pi (t-t_0)}{t_1-t_0}\right) \right)$,
we can integrate Eq.~\eqref{costH02} for $t\in[t_0,t_1]$ and find
\begin{equation}
\label{finalcostH0}
C_0=\hbar\left(\omega_1-\omega_0\right), 
\end{equation}
where the simplicity of the results stems from the choice of the state and the particular protocol.

\begin{figure}[t]
\includegraphics[width=0.9\columnwidth]{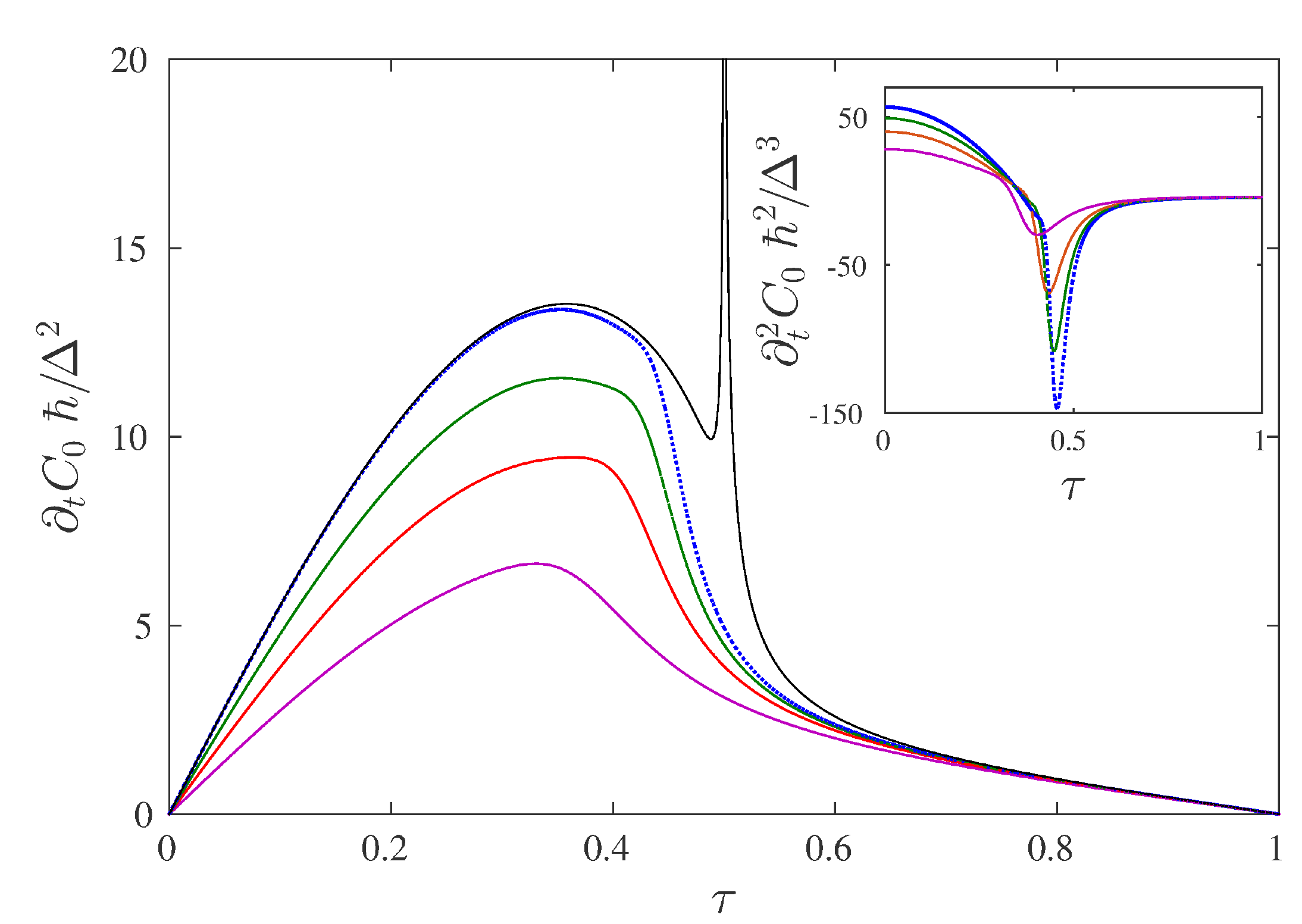}
\caption{(Color online) First derivative of the exigency, Eq.~\eqref{eq:C0}, when driving the groundstate of the LMG model through its quantum phase transition. The curves are for increasingly large system sizes from the bottom: $N=100$ (bottom, purple continuous curve), $N=200$ (red-dashed curve),  $N=300$ (green dot-dashed curve) and $N=400$ (blue-dotted curve). The top-most black thin continuous line is derived from the HP mapping with $N=400$. Inset: The second derivative of the exigency. The pronounced dip that approaches the critical point as system size increases is clearly visible. In both figures $\tau=(t-t_0)/(t_1-t_0)$ and $(t_1-t_0)=\Delta^{-1}\hbar$. } 
\label{fig:3} 
\end{figure}

\subsection{Exigency: Lipkin-Meshkov-Glick Model}
We now move to examining another critical many-body spin system, the Lipkin-Meshkov-Glick (LMG) model. A particularly interesting aspect is that it has infinite range interactions. It is therefore complementary to the short-range nearest neighbor Ising model previously studied. The LMG model can be solved analytically using the Holstein-Primakoff transformation~\cite{vidal1}. The exact form of the counterdiabatic term in the thermodynamic limit was calculated in Ref.~\cite{CampbellFazio2014}, shown to be non-local, and therefore the complexity associated with engineering exact counterdiabatic driving terms was linked to the closing energy gap near criticality. A means to circumvent the requirement to implement the full correction term was proposed, and it was found that significantly less resources were required to achieve effective adiabatic dynamics when far from criticality, while more refined correction terms were needed approaching the critical point. In what follows, we show through the use of the exigency this behavior in a more rigorous and quantitative manner. 

The LMG model in terms of collective spin operators, $\hat S_\alpha=\sum_i \hat \sigma_{\alpha}^i/2$ (where $\hat \sigma_\alpha$ are the usual Pauli operators) takes the form
\begin{equation}
\Hop_\text{LMG}(t)=-\frac{2\Delta}{N}\left( \hat S_x^2 + \gamma \hat S_y^2 \right) - 2g(t) \hat S_z. 
\end{equation}
For the time-independent case, the LMG model has a second order quantum phase transition when $g=\Delta$. To calculate the exigency we first notice that
\begin{equation}
\partial_t \Hop_\text{LMG} = -2\dot{g} \hat S_z.
\end{equation}
which allows us to directly evaluate the derivative of the exigency for pure states (see Appendix C) finding
\begin{equation}
\partial_t C_0 =2\sqrt{2}\vert\dot{g}\vert \sqrt{\text{Var}(\hat S_z)}.
\label{variance}
\end{equation}
It is immediately clear that $\partial_t\mathcal{C}_0=0$ when $\dot{g}=0$ or $\text{Var}(\hat S_z)=0$. In fact this result holds for any system where the driving is applied to a global field. In Fig.~\ref{fig:3} we show the behavior of Eq.~\eqref{variance}, evaluated when we drive the groundstate using the ramp, $g(t)/\Delta = \tfrac{3}{4} + \tfrac{1}{4}\left[1-\cos\left(\tfrac{\pi (t-t_0)}{t_1-t_0}\right)\right]$ for which $g=\Delta$ at $t=0.5(t_1+t_0)$. Clearly, the zero points at the start and end of the ramp are due to $\dot{g}=0$. We see that as the system size is increased the need to apply the counterdiabatic field grows, and does not appear to converge except for $t$ larger than $\frac{1}{2}(t_1+t_0)$. Furthermore, this need grows most significantly as we approach the critical point, implying that the cost associated with driving through the quantum phase transition diverges in the thermodynamic limit.  

Further insight can be found by exploiting the Holstein-Primakoff (HP) approximation that allows us to analytically treat the LMG (see Refs.~\cite{vidal1,CampbellFazio2014} for details). This mapping is exact in the thermodynamic limit, and provides an accurate approximation for suitably large $N$. Setting $\gamma=0$ for simplicity and dropping the explicit time dependence and defining $\tilde{g}=g/\Delta$ for brevity, through Eq.~\eqref{variance} we find that $\partial_t C_{0}$ is given by
\begin{equation}
\partial_t C_{0}=\begin{cases}
                 \label{limit}
                 2\vert \dot{g} \vert \sqrt{ \left(\tilde{g}\sinh^2(\alpha)+\frac{N\left( 1-\tilde{g}^2 \right)}{2} e^{\alpha}  \right) },   ~~&0<\tilde{g}<1 \\
                  2\vert \dot{g} \vert \sinh(\alpha),~~~&\tilde{g}>1,
\end{cases}
\end{equation}
where $\tanh(\alpha)=\frac{\tilde{g}^2}{2-\tilde{g}^2}$ for $0<\tilde{g}<1$, $\tanh(\alpha)=\frac{1}{2\tilde{g}-1}$ for $\tilde{g}>1$, and we have used the HP mapping (details provided in Appendix D). When $\tilde{g}>1$ we see the exigency is independent of the system size, however, for $0<\tilde{g}<1$ the need for a driving term scales with increasing $N$. In Fig.~\ref{fig:3} the solid black line corresponds to Eq.~\eqref{limit}. The approximation has excellent agreement with the numerics until we approach the critical point, where the mapping begins to break-down for any finite value of $N$. 

A final interesting point is the behavior of the second derivative of the exigency. In the inset of Fig.~\ref{fig:3} we see a divergence that is becoming increasingly more pronounced as we approach the critical point for systems tending towards the thermodynamic limit. We remark that this behavior is equivalent to that of other figures of merit which signal the emergence of critical behavior.

\section{Conclusions}
Enhancing the work output of quantum engines is key to designing future nano-technologies. External fields for counterdiabatic driving can increase the efficiency of thermal machines by increasing the work extracted in the unitary strokes. However the effectiveness of this method strongly depends on the cost of applying the external driving and on the duration of the process. While a quantitative evaluation is dependent on the particular experimental realization, in general, at longer time-scales the use of counterdiabatic driving is detrimental because the energy required to generate the field is larger than the energy gained from a quantum adiabatic evolution.          
The cost of applying counterdiabatic driving can be significantly reduced by choosing a selected form of driving which depends on the first measurement of energy, especially in systems close to a phase transition where the exact external driving is particularly costly for states more strongly affected by the transition. In future, optimizations comparing perfect against approximate counterdiabatic drivings and hybrid protocols that selects perfect or approximate drivings depending on the initial energy measurement could be implemented for considerable energy savings especially as quantum technologies are scaled-up. 
 
\acknowledgments{We are grateful to U. Bissbort, J. Gong and G. Xiao for fruitful discussions. This work is supported by SUTD start-up grant EPD2012-045, AcRF MOE Tier-II (project MOE2014-T2-2-119, WBS R-144-000-350-112), the John Templeton Foundation (grant ID 43467), and the EU Collaborative Project TherMiQ (Grant Agreement 618074). We thank COST Action MP1209 for partial support.}

\begin{appendix}

\section{Case of $8$ spins} \label{app:b}    

Here, we consider the illustrative case of $8$ spins for the Ising model explicitly. In this scenario, the Hilbert space comprises of $256$ states, $128$ each in the even or odd sector. In the even sector the possible values of the quasi-momentum are $k_e=\pm\pi/8$, $\pm 3\pi/8$, $\pm 5\pi/8$, $\pm 7\pi/8$, while in the odd sector $k_o=0$, $\pm\pi/4$, $\pm \pi/2$, $\pm 3\pi/4$, $\pi$. 

With a little computation, it becomes apparent that both even and odd sectors of the Hamiltonian are further divided into sub-blocks. This is due to the fact that in the basis spanned by the operators $c^{\dagger}_k$ acting on their vacuum $| v \ran$ ($c^{\dagger}_k |v\ran = 0$ for every $k$), only states differing by a pair of creation or destruction operators $\cop^{\dagger}_k\cop^{\dagger}_{-k}$ are coupled dynamically. 

For instance, one sub-block of the even sector, which we will refer to $S_A$, is spanned by the $16$ basis elements 
\begin{align}
S_A
\begin{cases}
 & |v\ran  \nonumber
\\ & \pop^{\dagger}_{1} |v\ran,\pop^{\dagger}_{3},|v\ran,\pop^{\dagger}_{5}|v\ran, \pop^{\dagger}_{7} |v\ran \nonumber
\\ & \pop^{\dagger}_{1}\pop^{\dagger}_{3} |v\ran, \pop^{\dagger}_{1}\pop^{\dagger}_{5} |v\ran, \pop^{\dagger}_{1}\pop^{\dagger}_{7} |v\ran, \pop^{\dagger}_{3}\pop^{\dagger}_{5} |v\ran, \pop^{\dagger}_{3}\pop^{\dagger}_{7} |v\ran, \pop^{\dagger}_{5}\pop^{\dagger}_{7} |v\ran
\\ & \pop^{\dagger}_{1}\pop^{\dagger}_{3}\pop^{\dagger}_{5} |v\ran, \pop^{\dagger}_{1}\pop^{\dagger}_{3}\pop^{\dagger}_{7} |v\ran, \pop^{\dagger}_{1}\pop^{\dagger}_{5}\pop^{\dagger}_{7} |v\ran, \pop^{\dagger}_{3}\pop^{\dagger}_{5}\pop^{\dagger}_{7} |v\ran 
\\ &  \pop^{\dagger}_{1}\pop^{\dagger}_{3}\pop^{\dagger}_{5}\pop^{\dagger}_{7} |v\ran
\end{cases}
\end{align}
where we have adopted the notation $\pop^{\dagger}_{j}=\cop^{\dagger}_{\frac{j\pi}{8}}\cop^{\dagger}_{-\frac{j\pi}{8}}$. 


Another block, $S_B$, is formed by single states that are completely uncoupled to any other state, (for example $\cop^{\dagger}_{\frac{\pi}8}\cop^{\dagger}_{\frac{3\pi}8}\cop^{\dagger}_{\frac{5\pi}8}\cop^{\dagger}_{\frac{7\pi}8} |v\ran$) because no pair $\pop^{\dagger}_{j}$ can be added to or be removed from it. By simple combinatorics, it is apparent that there are $16$ such states in the even sector. For example $\cop^{\dagger}_{\frac{\pi}8}\cop^{\dagger}_{-\frac{3\pi}8}\cop^{\dagger}_{-\frac{5\pi}8}\cop^{\dagger}_{\frac{7\pi}8} |v\ran$ is also one such state that can be obtained combinatorially.

A qualitatively intermediate scenario occurs in the sub-blocks $S_C$. For example one sub-block of $S_C$ could contain the states that are spanned by $\cop^{\dagger}_{\frac{\pi}8}\cop^{\dagger}_{{\frac{3\pi}8}} |v\ran$, $\cop^{\dagger}_{\frac{\pi}8}\cop^{\dagger}_{{\frac{3\pi}8}} \pop^{\dagger}_{5} |v\ran$, $\cop^{\dagger}_{\frac{\pi}8}\cop^{\dagger}_{{\frac{3\pi}8}} \pop^{\dagger}_{7} |v\ran$, $\cop^{\dagger}_{\frac{\pi}8}\cop^{\dagger}_{{\frac{3\pi}8}} \pop^{\dagger}_{5}\pop^{\dagger}_{7} |v\ran$. There are $24$ such sub-blocks in the even sector of the Hamiltonian that are of a similar structure (for example another group of $4$ coupled states is given by $\cop^{\dagger}_{-\frac{3\pi}8}\cop^{\dagger}_{{\frac{7\pi}8}} |v\ran$, $\cop^{\dagger}_{-\frac{3\pi}8}\pop^{\dagger}_{5}\cop^{\dagger}_{{\frac{7\pi}8}}  |v\ran$, $\pop^{\dagger}_{1} \cop^{\dagger}_{-\frac{3\pi}8}\cop^{\dagger}_{{\frac{7\pi}8}} |v\ran$, $ \pop^{\dagger}_{1}\cop^{\dagger}_{-\frac{3\pi}8}\pop^{\dagger}_{5}\cop^{\dagger}_{{\frac{7\pi}8}} |v\ran$). 

Now putting it all together, we find that the sub-block $S_A$ has $16$ states, all the sub-blocks of type $S_B$ have a total of $16$ states while there are a total of $96$ states in the sub-blocks of type $S_C$. The sum of all these states is indeed $128$ as expected. A similar scenario unravels in the odd sector. 

It is now clear that if the first energy measurement selects a state in any of the $S_B$ sub-blocks, no transitionless driving is needed because the states in $S_B$ are invariant during the time evolution. If instead the measurement selects an eigenstate of the sub-block $S_A$, then, in order to keep the evolution transitionless, it will be necessary to apply $\Hop_{I,_W}=\hat{P}_e\left[\sum_{k}f(k,t)\hat{\Psi}_k^{\dagger} \hat{\sigma}^y_k \hat{\Psi}_k \right]\hat{P}_e$ with $k$ given by all the possible $k_e$. Lastly, for a state in $S_C$, it would be sufficient to drive only two values of $k$ (which specific values of $k$ to be driven depends on which pairs are involved in the sub-block). For instance, in the example above,  $k=5\pi/8$ and $7\pi/8$ are needed.

\section{Time Evolution of a Diagonal Density Matrix in the Instantaneous Basis of the Hamiltonian}   \label{app:a}    

Here we furbish the details in deriving Eq.~\eqref{eq:evolution}. 
Assuming that the series expansion of $\rhom(t+dt)$ converges for sufficiently small $dt$, we write $$\rhom(t+dt)= \sum_n \frac{d^n\rhom(t)}{dt^n}\frac{dt^n}{n!}.$$ From here, using 
$$\frac{d\rhom(t)}{dt}=-\frac{\im}{\hbar}\left[\Hop_0(t),\rhom(t)\right]$$
we obtain the higher derivatives of $\rhom$, for instance,
\begin{align} 
\frac{d^2\rhom}{dt^2}=-\frac{1}{\hbar^2}\left[\Hop_0,\left[\Hop_0,\rhom\right]\right] -\frac{\im}{\hbar}\left[\frac{d\Hop_0}{dt},\rhom\right],  
\end{align}   
and 
\begin{align} 
\frac{d^3\rhom}{dt^3}&=\frac{\im}{\hbar^3}\left[\Hop_0,\left[\Hop_0,\left[\Hop_0,\rhom\right]\right]\right] -\frac{\im}{\hbar}\left[\frac{d^2\Hop_0}{dt^2},\rhom\right] \nonumber \\ 
& -\frac{1}{\hbar^2} \left\{ \left[\Hop_0,\left[\frac{d\Hop_0}{dt},\rhom\right]\right] + 2 \left[\frac{d\Hop_0}{dt},\left[\Hop_0,\rhom\right]\right]  \right\}.    
\end{align}  
Hence, it follows that if the $n-1$ derivatives of $\Hop_0$ commute with $\rhom$, then the lowest order correction in $dt$ to the time evolution of $\rhom$ will be given by Eq.~\eqref{eq:evolution} in the main text. 

\section{Exigency for the LMG model}
Here we detail the calculation to arrive at Eq.~\eqref{variance} in the main text. Assuming the state we wish to drive is $\rhom=\ket{\psi}\bra{\psi}$
\begin{equation*}
\begin{aligned}
\partial_t\mathcal{C}_0  &= \left\| \left[ \rhom,\partial_t\Hop_\text{LMG} \right] \right\| \\ 
                                           &= \left\| \rhom (-2\dot{g} \hat S_z) - (-2\dot{g} \hat S_z) \rhom \right\| \\
                                           &=2\vert\dot{g}\vert \sqrt{ \text{Tr}\left[ (\hat S_z\rhom - \rhom \hat S_z)(\rhom \hat S_z - \hat S_z\rhom) \right]  } \\
                                           &=2\vert\dot{g}\vert \sqrt{ \text{Tr}\left[\hat S_z\rhom^2 \hat S_z - \hat S_z\rhom \hat S_z\rhom - \rhom \hat S_z \rhom \hat S_z + \rhom \hat S_z^2\rhom \right]  } \\
                                           &=2\sqrt{2}\vert\dot{g}\vert  \sqrt{ \text{Tr}\left[ \rhom^2 \hat S_z^2 - \hat S_z\rhom \hat S_z\rhom \right] } \\
                                           &=2\sqrt{2}\vert\dot{g}\vert  \sqrt{  \bra{\psi}\hat S_z^2 \ket{\psi} - \bra{\psi} \hat S_z \ket{\psi}^2 } \\
                                           &=2\sqrt{2}\vert\dot{g}\vert \sqrt{\text{Var}(\hat S_z)}.
\end{aligned}
\end{equation*}

\section{Exigency for the LMG model: Holstein Primakoff approximation}
Here we outline the steps required in order to derive Eqs.~\eqref{limit}, with $\gamma=0$ for simplicity. This requires us to calculate Eq.~\eqref{variance} using the Holstein-Primakoff (HP) transformation. For suitably large $N$ we map the spin operators into the creation and annihilation operators $a$ and $a^{\dagger}$ of a harmonic oscillator 
\begin{eqnarray}
\hat S_x &=& \frac{\sqrt{N}}{2} \left(\hat a + \hat a^\dagger \right), \\
\hat S_z &=& \frac{N}{2}-\hat a^\dagger \hat a.
\end{eqnarray} 
Following the Supplementary material of~\cite{CampbellFazio2014}, when $\tilde{g}>1$ the HP transformation is always taken along $\hat S_z$. In order to map the LMG model to the harmonic oscillator we are required to perform a Bogoliubov transformation
\begin{eqnarray}
a&=\sinh\left(\frac{\alpha}{2}\right) b^\dagger + \cosh\left(\frac{\alpha}{2}\right) b,\\
a^\dagger&=\sinh\left(\frac{\alpha}{2}\right) b + \cosh\left(\frac{\alpha}{2}\right) b^\dagger,
\end{eqnarray}
with $\tanh \alpha = \tfrac{1}{2\tilde{g}-1}$. Therefore to calculate the required expectation value we must write $\hat a^\dagger \hat a$ in terms of $\hat b$ and $\hat b^\dagger$, and calculate the expectation value over the groundstate. It is readily found that the only term (other than constants) contributing is proportional to $\hat b^2 \hat b^{\dagger2}$. As such, we find that the variance of $\hat{S}_z$ is 
\begin{equation}
\text{Var}\left( \hat S_z \right) = \frac{1}{2} \sinh^2 \left(\alpha \right).
\end{equation}
Substituting into Eq.~\eqref{variance} we arrive at Eq.~\eqref{limit} (b)
\begin{equation}
\partial_t\mathcal{C}_0 =  2\vert \dot{g} \vert \sinh(\alpha).
\end{equation}

A similar, albeit more involved, calculation is required for $0<\tilde{g}<1$. In this case the direction along which the HP transformation must be taken changes with the value of $\tilde{g}$. In this case the operator of which we must calculate the variance of is
\begin{equation}
\hat S_z^\varphi  =  \hat S_z \cos\varphi  + \hat S_x\sin\varphi,
\end{equation}
with $\cos \varphi = \tilde{g}$. Once again, we express $\hat S_z^\varphi$ in terms of $\hat b$ and $\hat b^{\dagger}$ with $\tanh \alpha = \tfrac{\tilde{g}^2}{2-\tilde{g}^2}$, and calculate the expectation value over the groundstate. We now find that the only terms contributing are constants, and terms proportional to $\hat b \hat b^\dagger$ and $\hat b^2 \hat b^{\dagger2}$. After some manipulation we find
\begin{equation}
\begin{aligned}
\text{Var}\left( \hat S_z^\varphi \right) &= \left( \tfrac{\tilde{g}}{2} \sinh^2\left( \alpha \right) + \tfrac{N (1-\tilde{g}^2)}{4} \left[ \sinh\left( \tfrac{\alpha}{2} \right) +\cosh\left( \tfrac{\alpha}{2} \right) \right]^2 \right) \\
&= \left( \tfrac{\tilde{g}}{2} \sinh^2\left( \alpha \right) + \tfrac{N (1-\tilde{g}^2)}{4} e^\alpha \right).
\end{aligned}
\end{equation}
Substituting into Eq.~\eqref{variance} we arrive at Eq.~\eqref{limit} (a)
\begin{equation}
\partial_t\mathcal{C}_0 =  2\vert \dot{g} \vert \sqrt{ \left(\tilde{g}\sinh^2(\alpha)+\tfrac{N\left( 1-\tilde{g}^2 \right)}{2} e^{\alpha}  \right) }.
\end{equation}

\end{appendix}


\begin{thebibliography}{99}
\bibitem{notunique} Any driving $\Hop'_t=\Hop_t+\hat{P}_j\hat{O}_j\hat{P}_j$ where $\hat{O}_j$ is any operator, would also be a valid counterdiabatic driving field, however the one we chose is such that all its diagonal elements are $0$ and it has the smallest norm $\|\Hop_t\|$ \cite{DemirplackRice2008}.
\bibitem{MagneticVsElectric} If the driving Hamiltonian is proportional to a magnetic field $\Bv(t)$ (for example for a spin in a magnetic field) then the power needed to generate it will be proportional to $|\Bv|^2$ and thus to the square of the norm of the Hamiltonian $\|H_t\|^2$ (i.e. $n=2$). However if the driving is proportional to the square of an electric field $|\Ev(t)|^2$ (for example for optical lattices) then the power needed will be simply proportional to the norm of the Hamiltonian $\|H_t\|$ (i.e. $n=1$).
\bibitem{SameN} In this work we only consider external drivings produced by the same type of fields, i.e. only magnetic fields, such that we can use one unique exponent $n$ to compute the cost of the driving.
\bibitem{2levels} This is due to the fact that $\hat{P}_1+\hat{P}_2$ it is equal to the identity.
\bibitem{identnu} Here we consider that all $\nu_{_W,j,n}$ are identical.
\end{thebibliography}
\end{document}